\def\square{\,\hbox{\vrule\vbox{\hrule\phantom{N}\hrule}\vrule}\,}
 
 \parindent=0pt

\centerline{\bf Proofs of existence of local potentials for  trace-free symmetric 2-forms}
 
\centerline {\bf  using dimensionally dependent identities.}

\
 
 \centerline{\bf S. Brian Edgar. }

\centerline{Department of Mathematics, }

\centerline{Link\"{o}pings universitet,}

\centerline{Link\"{o}ping,}

\centerline{Sweden S-581 83.}
\smallskip
\centerline{ email: bredg@mai.liu.se} 

\

\

\

{\bf Abstract}

We  exploit  four-dimensional tensor identities to give a very simple  proof of the existence of a Lanczos potential for a Weyl  tensor in four dimensions with any signature, and to show that the potential  satisfies  a simple linear second order differential equation, e.g., a wave equation in Lorentz signature.  Furthermore, we  exploit higher dimensional tensor identities to obtain the analogous results  for  $(m,m)$-forms in $2m$ dimensions.

\

\

\

\

{\bf 1. Introduction.}

In analogy to the well-known vector potential $A_a$ for the electromagnetic field $F_{ab} = A_{[a;b]}$, Lanczos [1] proposed --- in four-dimensional curved spacetimes --- a $3$-tensor potential $L_{abc}$ for the Weyl tensor $C{^{ab}}{}_{cd}$, 
  $$\eqalign{ C^{ab}{}_{cd}  =  \   & L^{ab}{}_{[c;d]}
+L_{cd}{}^{[a;b]}  -  {}^*\!L^*{}^{ab}{}_{[c;d]}
-{}^*\!L^*_{cd}{}^{[a;b]}} 
\eqno(1) $$
where
$$L^{ab}{}_{c}=L^{[ab]}{}_{c} , \qquad L^{ab}{}_a=0, \qquad   L_{[abc]}=0\eqno(2a)
$$
and the double dual is defined for $L^{ab}{}_{[c;d]}$ in four dimensions by
$$ {}^*\!L^*{}^{ab}{}_{[c;d]}= {1\over 4} \eta^{abij}\eta_{cdpq}L_{ij}{}^{p:q} \  .
\eqno(2b)$$
It is easy to see that an equivalent version of (1) is
  $$\eqalign{ C^{ab}{}_{cd}  =  \  2 L^{ab}{}_{[c;d]}
+2L_{cd}{}^{[a;b]}  -2 
\delta^{[a}_{[c}\Bigl(L^{b]}{}^e{}_{d];e}+
L_{d]e}{}^{b];e}\Bigr) } 
 \eqno(1') $$
 Bambi and Caviglia [2] pointed out that Lanczos's proof was flawed, but gave an alternative proof of the existence of such a potential in four dimensions, emphasising that  the result applied to any {\it Weyl candidate tensor} $W_{abcd}$, i.e., any tensor with the index symmetries
 $$W^{ab}{}_{cd}=W^{[ab]}{}_{cd}=W^{ab}{}_{[cd]} = W_{cd}{}^{ab}, \qquad\qquad W_{a[bcd]}=0 \ .
 \eqno(3)$$ 
 Illge [3] has subsequently given a spinor proof of existence for Weyl candidates as part of a formal Cauchy problem analysis, and more recently a shorter spinor proof of existence has been given by Andersson and Edgar [4], which exploited a superpotential for Weyl candidates, i.e., a potential for the Lanczos potential. The proof in [2] was for any signature, while the spinor proofs in [3] and [4] are, by definition,   only valid in Lorentz signature. The key equation in the spinor proof in [4] was translated into tensors, from which  a tensor argument  was  constructed  which is valid for all signatures; however, since there was no direct tensor derivation of this key tensor equation, it is not possible to directly generalise this method in [4]  to $n> 4$ dimensions.  In this paper we will derive this key tensor equation in four dimensions {\it directly by tensor methods}, and so we will be able to explore the possibilities of  generalising to other dimensions.

 Some generalisations have been given in [3], [4], [5] for other spinors, and of course these results are also only valid for four-dimensional  spaces with Lorentz signature.

 It is important to note that although the electromagnetic potential  exists in arbitrary $n$ dimensions, the existence of the Lanczos potential  for Weyl candidates has only been confirmed in four dimensions. 
There exists a straightforward    $n$-dimensional  generalisation of the expression (1)
$$\eqalign{ W&^{ab}{}_{cd}  =  \cr &    L^{ab}{}_{[c;d]}
+L_{cd}{}^{[a;b]}  -  {1\over (n-4)!} \Bigl({}^*\!L^*{}^{i_1i_2  \ldots i_{n-4}ab}{}_{i_1i_2  \ldots i_{n-4}cd}
- {4\over (n-2)} {}^*\!L^*{}^{i_1i_2  \ldots i_{n-4}i_{n-3}[a}{}_{i_1i_2  \ldots i_{n-4}i_{n-3}[c}\delta^{d]}_{b]}\cr 
&\qquad\qquad\qquad\qquad\qquad\qquad  - {}^*\!L^*_{i_1i_2  \ldots i_{n-4}cd}{}^{i_1i_2  \ldots i_{n-4}ab} + {4\over (n-2)} {}^*\!L^*_{i_1i_2  \ldots i_{n-4}i_{n-3}[c}{}^{i_1i_2  \ldots i_{n-4}i_{n-3}[a}\delta_{b]}^{d]} \Bigr)} 
\eqno(4a) $$  
where the double dual of $L_{a_1a_2}{}^{[b_1;b_2]}$ is defined in $n$ dimensions by
$$ {}^*\!L^*{}^{a_3a_4  \ldots a_{n}}{}_{b_3 b_4  \ldots  b_{n}}= {1\over 4} \eta^{a_1a_2  \ldots a_{n}}\eta_{b_1b_2  \ldots  b_{n}}L_{a_1a_2}{}^{b_1;b_2} \  .
\eqno(4b)$$
The additional terms on the right hand side of (4a) compared to (1) ensure that $W^{ab}{}_{cd}$ is trace-free in all dimensions;  it should be noted that the double dual of $L_{a_1a_2}{}^{[b_1;b_2]}$ is  trace-free in four dimensions only, and even its multiple trace  ${}^*\!L^*{}^{i_1i_2  \ldots i_{n-4}ab}{}_{i_1i_2  \ldots i_{n-4}cd}$ is not trace-free in $n>4$ dimensions.  
More concisely we can rearrange (4) as
  $$\eqalign{ W^{ab}{}_{cd}  =  \  2 L^{ab}{}_{[c;d]}
+2L_{cd}{}^{[a;b]}  -{4\over (n-2)} 
\delta^{[a}_{[c}\Bigl(L^{b]}{}^e{}_{d];e}+
L_{d]e}{}^{b];e}\Bigr) } 
 \eqno(4') $$
where   $L_{abc}$ and $W^{ab}{}_{cd}$ retain  the respective properties (2a) and (3).   (See [6] for the properties of double duals.) However, we emphasise that
 it has been shown that such a Lanczos potential {\it cannot exist}  in general, in dimensions $n>4$ [7]. 
 This result in [7] seems to contradict some comments in [2], where it is suggested that there does exist, in general for arbitrary $W^{ab}{}_{cd}$,  a Lanczos potential satisfying $(4')$ in  five  dimensions. (This conclusion is also stated to be true for {\it six} dimensions in [2], but it is easy to see that there is a simple computational error in the last step of the argument, which when corrected excludes six dimensions.) So there must be some unease about the validity of the (long and complicated) tensor proof in [2] for {\it four} dimensions, since the same arguments as are used in the four dimensional proof seem to be used to prove existence in five dimensions in [2] --- a result which we now know not to be true.  So it would be preferable to have an alternative proof of existence in four dimensions using tensors,  valid in all signatures.
 \smallskip
 More than thirty years ago, Lovelock [8] drew attention to the significance of {\it dimensionally dependent tensor identities} (DDIs), and exploited them to unify a number of apparently unrelated results. Recently such identities have been shown to be a useful and powerful tool which have been used to generalise spinor proofs to tensor proofs (valid in four-dimensional spaces of all signatures), and also to generate, in higher dimensions,  new results analogous to familiar  results in four dimensions [9], [10], [11], [12].
  \smallskip
 In this paper we will exploit two four-dimensional DDIs to give a very simple   proof of the local existence of a Lanczos potential of a Weyl candidate tensor in any signature; the argument used is a tensor version of the spinor proof in [4]. Furthermore,  although we now know from [7] that there is no direct higher dimensional analogue of (1) via (4) for all Weyl candidates,  there is the possibility of some other analogue to the Lanczos potential for {\it  other types of tensors} in higher dimensions.  By considering the  higher dimensional DDIs analogous to the four dimensional ones, we will establish local existence of potentials for symmetric trace-free $(m,m)$-forms in $2m$ dimensions.

 One of the most interesting properties of the Lanczos potential for the Weyl curvature tensor is the fact that in vacuum it satisfies the very simple Illge wave equation  in four dimensional spacetime, $\square L_{abc}=0$  [3]; using DDIs we shall investigate analogous results in other signatures and  in higher dimensions.

 \
 
 \
 
 {\bf 2. Potentials in Four Dimensions.}

  For a trace-free symmetric   $(2,2)$-form $V^{ab}{}_{cd}=
V^{[ab]}{}_{cd}=V^{ab}{}_{[cd]}=V_{cd}{}^{ab}, \ V^{ab}{}_{ad}=0$,   there exists   a {\it
four-dimensional} DDI [8,9]
$$V^{[ab}{}_{[cd}\ \delta^{e]}_{f]}\equiv0
 \eqno(5)$$
which, when differentiated twice with $\nabla_e \nabla^f$, yields
$$ \eqalign{V^{ab}{}_{cd}{}^{;e}{}_{e} &  = 2
V^{[a|e|}{}_{cd}{}^{;b]}{}_e+2 V^{ab}{}_{[c|e|}{}^{;e}{}_{d]} -4
 V^{[a|e}{}_{[c|f}{}^{;f|}{}_{e|}\delta^{b]}_{d]} \cr & 
 = 2
V^{[a|e|}{}_{cd}{}_{;e}{}^{b]}+2 V^{ab}{}_{[c|e|}{}^{;e}{}_{d]}   -4
 V^{[a|e}{}_{[c|f}{}^{;f|}{}_{e|}\delta^{b]}_{d]} \cr & \qquad \qquad
 - R^{abie}V_{ie}{}_{cd} - 2R^{[b}{}_e{}_{c}{}_iV^{a]e}{}^{i}{}_{d}-2R^{[b}{}_e{}_{d}{}_iV^{a]e}{}_c{}^{i}+2 R^{[b}{}_i V^{a]i}{}_{cd}   
 }\eqno(6)$$

Defining
$$H^{ab}{}_c= V^{ab}{}_{ce}{}^{;e}=  V_{ce}{}^{ab;e}
 \eqno(7)$$
where we have exploited the symmetry of the $(2,2)$-form.  It follows that 
$$H^{ab}{}_a=0 \ , \eqno(8)$$
and we obtain 
$$ \eqalign{ \nabla^2 V^{ab}{}_{cd}+  R^{abie}V_{ie}{}_{cd} + & 4R_e{}^{[a}{}_{[c}{}_{|i|}V^{b]e}{}^{i}{}_{d]} -  R^{ei}{}^{f[a}V_{ei}{}_{f[c}\delta_{d]}^{b]}  + R_{ei}{}_{f[c}
  V{}^{ei}{}^{f[a}\delta_{d]}^{b]}    +2 R^{[a}{}_i V^{b]i}{}_{cd}   \cr &  \qquad \quad = 2
H_{cd}{}^{[a;b]}+2 H^{ab}{}_{[c;d]}-2
H^{[a|e|}{}_{[c}{}_{;|e|}\delta^{b]}_{d]} -2
 H_{[c|f|}{}^{[a}{}^{;|f|}\delta^{b]}_{d]} 
}\eqno(9)$$
where it is easily confirmed that the right hand side  has the structure of a trace-free symmetric $(2,2)$-form.
On the otherhand, although the left hand is easily confirmed to be trace-free, it  is not obviously  a {\it symmetric} $(2,2)$-form; in order for the left hand side to have that structure, it would be necessary that
  $$ \eqalign{  R^{abij}V_{ij}{}_{cd} - R_{cdij}V^{ij}{}^{ab}    -  2 R^{ei}{}^{f[a}V_{ei}{}_{f[c}\delta_{d]}^{b]}  + 2R_{ei}{}_{f[c}
  V{}^{ei}{}^{f[a}\delta_{d]}^{b]}    +2 R^{[a}{}_i V^{b]i}{}_{cd} -2  R_{[c}{}^i V_{d]i}{}^{ab}  =0}\eqno(10)$$
  
 But precisely this DDI can be constructed  from the DDI (5), via
$$R^{ej}{}_{id}V^{[ab}{}_{[ce}\ \delta^{i]}_{j]}\ =0
 \eqno(11)$$ 

\smallskip

Using (10) to rearrange   (9), followed by the decomposition of the Riemann tensor into its Weyl and Ricci parts, gives
$$ \eqalign{ \nabla^2 V^{ab}{}_{cd}+  & {1\over 2}(C^{abij}V_{ij}{}_{cd} + C_{cdij}V^{ij}{}^{ab}) +  4C_e{}^{[a}{}_{[c}{}_{|i|}V^{b]e}{}^{i}{}_{d]}    +{R\over 2} V^{ab}{}_{cd} \cr &  \qquad \quad = 2
H_{cd}{}^{[a;b]}+2 H^{ab}{}_{[c;d]}-2
H^{[a|e|}{}_{[c}{}_{;|e|}\delta^{b]}_{d]} -2
 H_{[c|f|}{}^{[a}{}^{;|f|}\delta^{b]}_{d]} 
}\eqno(9')$$
which not only has the explicit structure of a trace-free symmetric $(2,2)$-form on both sides as required, but has also a considerably simpler left hand side; note the absence of Ricci scalar terms.

Now consider {\it any} trace-free  
symmetric  $(2,2)$-form $U^{ab}{}_{cd}=
U^{[ab]}{}_{cd}=U^{ab}{}_{[cd]}=U_{cd}{}^{ab}, \ U^{ab}{}_{ad}=0$. We can
always find a trace-free  
symmetric $(2,2)$-form 'superpotential' $V^{ab}{}_{cd}$ locally   for $U^{ab}{}_{cd}$ by appealing to the Cauchy-Kovalevskaya theorem which guarantees a local analytic solution of
$$ \eqalign{ \nabla^2 V^{ab}{}_{cd}+  & {1\over 2}(C^{abij}V_{ij}{}_{cd} + C_{cdij}V^{ij}{}^{ab}) +  4C_e{}^{[a}{}_{[c}{}_{|i|}V^{b]e}{}^{i}{}_{d]}    +{R\over 2} V^{ab}{}_{cd}    = U^{ab}{}_{cd}
}\eqno(12)$$
in a given background space. From the superpotential $V^{ab}{}_{cd}$ we can then construct 
a potential $H^{ab}{}_{c}$ from (7), and obtain the following result:
  
  \smallskip
  
{\bf Theorem 1.}  {\it In four dimensions, the trace-free symmetric  $(2,2)$-form $ U^{ab}{}_{cd}$ has a trace-free $(2,1)$-form potential $ H^{ab}{}_{c}$, given by
$$ U^{ab}{}_{cd} = 2
H_{cd}{}^{[a;b]}+2 H^{ab}{}_{[c;d]}-2
H^{[a|e|}{}_{[c}{}_{;|e|}\delta^{b]}_{d]} -2
 H_{[c|f|}{}^{[a}{}^{;|f|}\delta^{b]}_{d]}  
\ .
 \eqno(13)$$
Furthermore this potential $H_{abc}$ satisfies
$$\eqalign{
&
  \nabla^2 H^{ab}{}_{c} 
+(H_{c}{}^{e}{}^{[a}   -H^{[a}{}^{|e|}{}_c){}_{;e}{}^{b]}
  -H^{ab}{}_{e}{}^{;e}{}_{c}  -\delta_{c}^{[a}H^{b]}{}^{e}{}_{d}{}^{;d}{}_{e}
 \cr& \qquad\
 -{3\over 2}H_{[cde]}R^{de}{}^{ab}
  +3H^{[bde]}R_c{}^a{}_{ed}\cr& \qquad\qquad  -2(H_{ce}{}^{[a}- H^{[a|e|}{}_c)R^{b]e}
  +2 H^{ab}{}_eR^e{}_c +2 H^{f[a}{}_e\delta^{b]}_cR^e{}_f  -{1\over 2}H^{ab}{}_c R=U^{ab}{}_{cd}{}^{;d}.
}\eqno(14)$$}
  \smallskip

To obtain (14) we take the divergence of (13) and rearrange into the form
$$\eqalign{
&
  \nabla^2 H^{ab}{}_{c} 
+(H_{c}{}^{e}{}^{[a}   -H^{[a}{}^{|e|}{}_c){}_{;e}{}^{b]}
  -H^{ab}{}_{e}{}^{;e}{}_{c}  -\delta_{c}^{[a}H^{b]}{}^{e}{}_{d}{}^{;d}{}_{e}
 \cr& \qquad\
+2H^{e[b}{}_{d}R^{a]d}{}_{ec}
  +{1\over 2}(H_{dec}-3H_{[cde]})R^{de}{}^{ab}
  +H^{ef}{}_{d}\delta_{c}^{[a}R^{b]d}{}_{ef}+3H^{[bde]}R_c{}^a{}_{ed}
\cr& \qquad\qquad\qquad\qquad\qquad\qquad\qquad\qquad\qquad\qquad\qquad\qquad
  +2H_{ce}{}^{[b}R^{a]e}
  +H^{ab}{}_{e}R_{c}{}^{e}
 =U^{ab}{}_{cd}{}^{;d}.
}\eqno(15)$$
We now note that due to the existence of the {\it four-dimensional} DDI [9]
$$H^{[ab}{}_{[c}\delta_{ef]}^{gh]} \equiv 0
\eqno(16)$$
we obtain 
$$\eqalign{0\equiv & 12R_{gh}{}^{ef} H^{[ab}{}_{[c}\delta_{ef]}^{gh]} = 4H^{e[b}{}_{d}R^{a]d}{}_{e}{}_{c}
  +H^{de}{}_{c}R_{de}{}^{ab}
  +2H^{ef}{}_{d}\delta_{c}^{[a}R^{b]d}{}_{ef} \cr& \qquad\qquad\qquad\qquad\qquad\qquad\qquad  - 2H^{ab}{}_eR^e{}_c+4H^{e[a}{}_cR^{b]}{}_e - 4 H^{f[a}{}_e\delta^{b]}_cR^e{}_f +H^{ab}{}_c R} 
\eqno(17)$$
whose substitution in (15) results in  (14). (Alternatively, we could exploit the four-dimensional identity $C_{[ab}{}^{[cd}\delta^{e]}_{f]}\equiv  0$ [13], [14]).

The fact that (14) is significantly simpler than (15) is apparent  when we restrict the superpotential by the additional symmetry of the type of the 'first Bianchi identity'  $V_{[abc]d}=0$,  leading to  $H_{[abc]}=0$ and  the four-dimensional version of (4) for the Lanczos potential $L_{abc}(\equiv H_{abc})$ of a Weyl candidate tensor $W^{ab}{}_{cd}(\equiv U^{ab}{}_{cd})$, or in particular,  $(1')$ for  the Weyl tensor $C^{ab}{}_{cd}$. 

 \smallskip
  
{\bf Corollary 1:1.}  {\it In four dimensions, the Weyl tensor    $ C^{ab}{}_{cd}$ with the properties (3) has a  potential $ L^{ab}{}_{c}$ with the properties (2a), given by
$$ C^{ab}{}_{cd} = 2
L_{cd}{}^{[a;b]}+2 L^{ab}{}_{[c;d]}-2
L^{[a|e|}{}_{[c}{}_{;|e|}\delta^{b]}_{d]} -2
 L_{[c|f|}{}^{[a}{}^{;|f|}\delta^{b]}_{d]}  
\ .
 \eqno(18)$$
Furthermore this potential $L_{abc}$ satisfies
$$\eqalign{
&
  \nabla^2 L^{ab}{}_{c} 
+L^{[a}{}_{c}{}_e  {}^{;|e|}{}^{b]}
  -L^{ab}{}_{e}{}^{;e}{}_{c}  -\delta_{c}^{[a}L^{b]}{}^{e}{}_{d}{}^{;d}{}_{e}
  +2L^{[a}{}_{ce}{}R^{b]e}
  +2 L^{ab}{}_eR^e{}_c +2 L^{f[a}{}_e\delta^{b]}_cR^e{}_f  -{1\over 2}L^{ab}{}_c R   
\cr & \qquad \qquad \qquad \qquad \qquad \qquad \qquad \qquad \qquad\qquad \qquad \qquad \qquad  \qquad = C^{ab}{}_{cd}{}^{;d}  \cr & \qquad \qquad \qquad \qquad \qquad \qquad \qquad \qquad \qquad\qquad \qquad \qquad \qquad  \qquad = - R_c{}^{[a;b]}-\delta_c^{[a}R^{,b]}/6 }\eqno(19)$$}
  \smallskip
So,  the simplification from (15) to (14) is especially significant when we are investigating the Lanczos  $L_{abc}$ potential of a Weyl candidate, and in particular of the Weyl tensor $C_{abcd}$; we note that the DDI (17) leads to    the absence of all Weyl tensor terms in (19). 
 The absence of the Weyl tensor terms in (19) was completely unsuspected in the original tensor calculations [1] and in a number of subsequent papers, e.g., [15], [16].  There it was assumed that there existed, on the left hand side of (14), explicit terms involving the Weyl tensor, and that the substitution of the definition of the Lanczos potential  into those terms would lead to  a non-linear  differential equation for $L_{abc}$; indeed there were attempts to deal with this equation  by taking linear approximations [1], [16]. The spinor derivations of Illge [3] revealed the remarkable simplification of the absence of Weyl tensor terms meaning that there were no non-linear terms for $L_{abc}$;  this was subsequently confirmed by the tensor derivation, similar to above, for all signatures in four dimensions in [13,14].

\

\

  {\bf 3. Differential gauge in four dimensions.} 

For the Lanczos potential, it is well known  [1-6,14]  that, in general, 
there exists a 'differential gauge' term $L_{ab}{}^c{}_{;c}$ which  can be given any value; 'the Lanczos differential gauge'  is when this choice is zero. In particular, the equation (19) would become even simpler  if the Lanczos differential gauge choice could be made.
In the approach in the previous section we do not have this gauge freedom, since we have fixed the potential $H_{abc}$  in terms of the superpotential  $V_{abcd}$. In this section we shall show how we can modify the approach in the previous section to include this differential gauge freedom.  

We begin as in the last section with the trace-free symmetric   $(2,2)$-form $V^{ab}{}_{cd}=
V^{[ab]}{}_{cd}=V^{ab}{}_{[cd]}=V_{cd}{}^{ab}, \ V^{ab}{}_{ad}=0$,  and since we wish to concentrate on Weyl tensors, we shall include the additional condition $V_{[abc]d}=0$ from the beginning. The DDI (5) leads to the differential equation (6).

Next we define
$$L^{ab}{}_c= V^{ab}{}_{ce}{}^{;e} +{1\over 2} (F^{ab}{}_{;c}-F_{c}{}^{[a;b]})+{1\over 2} \delta_{c}^{[a}F^{b]}{}^d{}_{;d}
 \eqno(7^*)$$
  where $F_{ab}$ is an arbitrary $2$-form, and the last two terms ensure that $L^{ab}{}_c$ retains the   properties (2a).
  Substitution of $(7^* )$ into (6) leads to 
  $$ \eqalign{ \nabla^2 V^{ab}{}_{cd}+  & {1\over 2}(C^{abij}V_{ij}{}_{cd} + C_{cdij}V^{ij}{}^{ab}) +  4C_e{}^{[a}{}_{[c}{}_{|i|}V^{b]e}{}^{i}{}_{d]}    +{R\over 2} V^{ab}{}_{cd} +{3\over 2} (C^{ab}{}_{[c}{}^eF_{d]e} +C_{cd}{}^{[a}{}_eF^{b]}{}^{e})\cr &  \qquad \quad = C^{ab}{}_{cd}
}\eqno(12^*)$$
 where 
 $$C^{ab}{}_{cd}= 2
L_{cd}{}^{[a;b]}+2 L^{ab}{}_{[c;d]}-2
L^{[a|e|}{}_{[c}{}_{;|e|}\delta^{b]}_{d]} -2
 L_{[c|f|}{}^{[a}{}^{;|f|}\delta^{b]}_{d]} \eqno(13^*)
  $$
  and the argument continues by exploiting the DDI (11) as in the last section. The presence of the arbitrary $2$-form $F_{ab}$ in our definition $(7^* )$ and in the equation $(12^* )$ does not effect either equations (18) or (19), and so we are free to choose $F_{ab}$ as we wish. 
  
By taking the divergence of  $(7^* )$ we find 
  $$\nabla^2F_{ab}+4C^{def}{}_{[a}V_{b]def}+{1\over 2}C_{ab}{}^{de}F_{de}-{R\over 6}V_{ab}= 2L_{abc}{}^{;c}
  \eqno(20)$$
 and so instead of finding our superpotential $V_{abcd}$ as a   local solution of (12) as in the last section,  we can instead consider the coupled pair of equations $(12^* )$ and $(20 )$ which guarantee locally  solutions $V_{abcd}$ and   $F_{ab}$ given the Weyl tensor $C_{abcd}$ and  the differential gauge $L_{abc}{}^{;c}$; in particular we can choose the Lanczos differential gauge, $L_{abc}{}^{;c}=0$.
 
 Hence we can modify Corollary 1.1 to:
 \smallskip
{\bf Corollary 1:2.}  {\it In four dimensions, the Weyl tensor    $ C^{ab}{}_{cd}$ with the properties (3) has a  potential $ L^{ab}{}_{c}$ with the properties (2a), given by
$$ C^{ab}{}_{cd} = 2
L_{cd}{}^{[a;b]}+2 L^{ab}{}_{[c;d]}-2
L^{[a|e|}{}_{[c}{}_{;|e|}\delta^{b]}_{d]} -2
 L_{[c|f|}{}^{[a}{}^{;|f|}\delta^{b]}_{d]}  
\ .
 \eqno(18^*)$$
There exists the freedom to choose the differential gauge $L_{abc}{}^{;c}$; in particular for the Lanczos differential gauge $L_{abc}{}^{;c}=0$,  this potential $L_{abc}$ satisfies
$$\eqalign{
&
  \nabla^2 L^{ab}{}_{c} 
  +2L^{[a}{}_{ce}{}R^{b]e}
  +2 L^{ab}{}_eR^e{}_c +2 L^{f[a}{}_e\delta^{b]}_cR^e{}_f  -{1\over 2}L^{ab}{}_c R   
 = C^{ab}{}_{cd}{}^{;d}  \cr & \qquad \qquad \qquad \qquad \qquad \qquad \qquad \qquad \qquad\qquad \qquad \qquad \quad \  = - R_c{}^{[a;b]}-\delta_c^{[a}R^{,b]}/6 }\eqno(19^*)$$}
  \smallskip

\

\
 
  {\bf 4. Potentials in Even Dimensions.} 
  
An obvious generalisation involves the trace-free symmetric $(m,m)$-forms 
 $$V^{a_1a_2\ldots a_m}{}_{b_1b_2\ldots b_{m}}=
V^{[a_1a_2\ldots a_m]}{}_{b_1b_2\ldots b_{m}}=V^{a_1a_2\ldots a_m}{}_{[b_1b_2\ldots b_{m}]}=V_{b_1b_2\ldots b_m}{}^{a_1a_2\ldots a_m} \eqno(21a)$$
 $$ V^{a_1a_2\ldots a_m}{}_{a_1b_2\ldots b_m}=0,  \eqno(21b)$$
 in even dimensions $n=2m$.  By exploiting the {\it $2m$-dimensional} DDI [8,9],
$$
  V^{[a_1a_2\ldots a_m}{}_{[b_1b_2\ldots b_{m}}\delta^{e]}_{f]}\equiv 0
 \eqno(22)$$
and following the same arguments as in the four-dimensional case we obtain the result: 

\smallskip
{\bf Theorem 2.} {\it  In even $n=2m$ dimensions,  the  trace-free symmetric  $(m,m)$-form $ U^{a_1a_2\ldots a_m}{}_{b_1b_2\ldots b_{m}}$ has a trace-free $(m,m-1)$-form potential $ H^{a_1a_2\ldots a_m}{}_{b_1b_2\ldots b_{m-1}}$, given by
$$ \eqalign{U^{a_1a_2\ldots a_m}{}_{b_1b_2\ldots b_{m}} & = m
H_{b_1b_2\ldots  b_{m}}{}^{[a_1a_2\ldots a_{m-1};a_m]}+ mH^{a_1a_2\ldots a_m}{}_{[b_1b_2\ldots b_{m-1} ;b_{m}]}
\cr & \quad-{m^2\over 2}
H^{[a_1a_2\ldots a_{m-1}|e|}{}_{[b_1 b_2\ldots b_{m-1} ;|e|} \delta^{a_{m}]}_{b_{m}]}
-{m^2\over 2}
H_{[b_1 b_2\ldots b_{m-1}|e|}{}^{[a_1a_2\ldots a_{m-1};|e|} \delta^{a_m]}_{b_m] }} \eqno(23)$$}

\smallskip

Clearly a second order equation for the potential can be obtained by taking the divergence of (23), and it is of interest to know whether  analogous simplifications as occured in the four dimensional case also occur in  higher dimensions. 

As an example we  will examine the   trace-free symmetric  $(3,3)$-form 
$$ \eqalign{U^{abc}{}_{def} & = 3
H_{def}{}^{[ab;c]}+ 3H^{abc}{}_{[de ;f]}
-{9\over 2}
H^{[ab|i|}{}_{[de ;|i|} \delta^{c]}_{f]}
-{9\over 2}
H_{[de|i|}{}^{[ab;|i|} \delta^{c]}_{f] }} \eqno(24)$$
in six dimensions.
We obtained some simplifications in the four dimensional case, when we added the additional symmetry property $V_{[abc]d}=0$ to the superpotential leading to the additional property $L_{[abc]}=0$  on the potential for the Weyl candidate, which of course satisfies $W_{[abc]d}=0$ ; in an analogous manner we can choose,  in this six dimensional case, 
 $$V_{[abcde]f}=0 \qquad  \hbox{leading to}  \qquad H_{[abcde]}=0  \qquad \hbox{and  hence }  \qquad U_{[abcde]f}=0\eqno(25)$$
Again, in analogy with the four dimensional case, we can introduce the differential gauge freedom by redefining 
$$H^{abc}{}_{de}= V^{abc}{}_{def}{}^{;f} +{1\over 2} \bigl(F^{abc}{}_{[d;e]}-F_{de}{}^{[ab;c]}\bigr)-{1\over 3} \delta_{[d}^{[a}\bigl(F^{bc]i}{}_{e];i}-F_{e]i}{}^{bc];i}\bigr)
 \eqno(26)$$
  where $F^{abc}{}_d$ is an arbitrary $(3,1)$-form, and the last two terms ensure that $H^{abc}{}_{de}$ retains  the trace-free property as well as the  property in (25). By an appropriate choice of $F^{abc}{}_d$, we can choose the differential gauge 
  $$H^{abc}{}_{de}{}^{;e}=0\eqno(27)$$

Whenever we take the divergence of (24) we obtain 
$$ \eqalign{
\nabla^2 H^{abc}{}_{de} + & 2 H^{abc}{}_{[e|f|;d]}{}^f+3H_{def}{}^{[ab;c]f}\cr & -{3\over 2}H^{[ab|i|}{}_{de;i}{}^{c]} -3H^{[ab|i}{}_{[e|f;i|}{}^{f|} \delta^{c]}_{d]}-{3\over 2}H_{dei}{}^{[ab;|i|c]}
-3H_{f[d|i|}{}^{[ab;|i}{}^{f|} \delta^{c]}_{e] }=U^{abc}{}_{def}{}^{;f}
}\eqno(28)
$$
which we can rearrange to
$$ \eqalign{
\nabla^2 H^{abc}{}_{de} + & 2 H^{abc}{}_{[e|f|}{}^{;f}{}_{d]}+3H_{def}{}^{[ab;|f|c]} \cr & -{3\over 2}H^{[ab|i|}{}_{de;i}{}^{c]} -3H^{[ab|i}{}_{[e|f;i|}{}^{f|} \delta^{c]}_{d]}-{3\over 2}H_{dei}{}^{[ab;|i|c]}
+\Bigl\{R\otimes H\Bigr\}^{abc}{}_{de}=U^{abc}{}_{def}{}^{;f}
}\eqno(29)
$$
where $\Bigl\{R\otimes H\Bigr\}^{abc}{}_{de}$ represents the rather long collection of terms involving products of the Riemann tensor and the potential tensor. Next making use of the additional symmetry property (25), together with the differential gauge choice (27),  equation (29) simplifies to 
$$ \eqalign{
\nabla^2 H^{abc}{}_{de} 
+ \Bigl\{R\otimes H\Bigr\}^{abc}{}_{de}=U^{abc}{}_{def}{}^{;f}
}\eqno(30)
$$

A further rearrangement of (30) is possible, where, in analogy with the four dimensional case, we can exploit the six dimensional DDI
$$
R^{ij}{}_{kl} H^{[abc}{}_{[de}\delta^{kl]}_{ij]}\equiv0\eqno(31)
$$

Without further  detailed calculations it is not possible to determine the extent of these simplifications, and in particular  whether all the Weyl tensor terms disappear, as happened in the four dimensional case.
Even for the six dimensional case, the calculations are considerably longer than in the four dimensional case, but it is emphasised that the same approach can be used in all $2m$ dimensions.

Unfortunately, unlike in the four dimensional case, there do not seem obvious geometric or physical candidates ($(m,m)$-forms with $m>2$) to which  these higher dimensional results can be applied. 

\

\

{\bf 4. Discussion.}

First we make some general comments on the uses of DDIs:

$\bullet$ \  A DDI for $n$ dimensions is constructed by antisymmetrising over $n+1$ indices  and so, in general, is valid for $n$ dimensions {\it and lower}. This applies to the results also in this paper; however for these lower dimension the results are trivial since from [9] it is known that the  trace-free  $(p,q)$-form  $V^{a_1a_2\ldots a_p}{}_{b_1b_2\ldots b_q}$ is identically zero in dimensions $n< (p+q)$.

\smallskip

$\bullet$ \   In our four-dimensional investigations in Section 2, we exploited the  four-dimensional DDI (5) on two occasions, and the DDI (16) on one occasion; in the analogous spinor investigations [4] there was no need to supply this input explicitly since it is inbuilt into the spinor formalism. The advantage of having the alternative tensor investigation is that it suggests how and where generalisations can be made to $n>4$ dimensions, and  signposts where simplifications might be expected. In $n>4$ dimensions  we exploited the analogous $2m$-dimensional DDI  (22) in the same two stages of the argument with  analogous results,  and we also noted where we could  use the other  analogous DDI (31), in the six dimensional case. The full details of these calculations for six dimensions, and more generally $2m$ dimensions will be reported elsewhere.

\smallskip

$\bullet$ \  Familiar and useful results in four dimensions can be generalised to higher dimensions
 by the use of the analogous dimensionally dependent identities in higher dimensions. But the type of tensor to which the familiar four-dimensional results apply is also generalised, and so the higher dimensional generalisations often apply to less common types of tensors. So, for instance, there is no possibility of using dimensionally dependent identities to obtain potentials for the Weyl tensor in dimensions higher than four.

 \smallskip

$\bullet$ \  It might be suspected that more general identities than (5), (16), (22), (31)  could be obtained for forms without the trace-free restriction; this is not possible [9].  So, for instance, there is no possibility of using dimensionally dependent identities to obtain potentials for the {\it Riemann} tensor in four dimensions. Furthermore no other identities can be constructed by using the traces of the fundamental dimensionally dependent identities such as (5), (16),  (22), (31) because these collapse to be trivially zero.

\smallskip

We next make some comments specifically concerned with applications to Weyl and Riemann tensors:

\smallskip
$\bullet$ \  We have seen that  there are properties of the Weyl tensor which are very different in four dimensions than in other dimensions; but this has nothing to do with its differential structure, since these special properties in four dimensions are algebraic, and arise  because  {\it any}  symmetric trace-free $(2,2)$-form has a very special role in four dimensions. 

\smallskip
$\bullet$ \  It is clear that the existence of a Lanczos-type  potential for a tensor --- which requires no differential conditions on the tensor --- is a very different type of result from the existence of  the electromagnetic potential --- which follows from differential conditions via Poincar\'e's Lemma. However it is interesting to note that the existence of  the electromagnetic potential  in {\it four} dimensions can be deduced via a Lanczos-type potential.
The additional results for the existence of other spinor potentials in [4] includes for instance a spinor proof for the existence for the spinor potential for the electromagnetic field; this is a {\it complex} potential, which  can be reduced to a real potential by the first Maxwell equation $F_{[ab;c]}=0$ [3,4].

\smallskip

From the point of view of general relativity, we note:
\smallskip
$\bullet$ We can appeal to  stronger existence theorems [17] when we specialise to  spaces with Lorentz signature, and the second order differential equations  become wave equations.

\smallskip
$\bullet$ If (in Lorentz signature) we replace the Ricci tensor with the energy-momentum tensor   via Einstein's equations in  (19) or $(19^*)$ we find that  we have   to solve for the Lanczos potential from
  a wave equation whose other terms contain only the energy momentum tensor. This  linear wave equation for $L_{abc}$, which carries much of the information of Einstein's equations, is considerably simpler than the non-linear wave equation for $C_{abcd}$, which is being used in a number of applications.

\smallskip

Finally we emphasise that, although it is not so obvious,  it is possible to apply this approach to $(p,q)$-forms even when $p\ne q$, as we will report in a subsequent paper. 

\

\

{\bf Acknowledgements.}

Thanks to Jos\'{e} Senovilla for  comments,  discussions and suggestions.

 \
 
 \
 
{\bf References}.

1. Lanczos, C. (1962). The splitting of the Riemann tensor. {\it  Rev. Mod. Phys.}, {\bf 
34}, 379.

2. Bampi, F., and Caviglia, G. (1983). Third-order tensor potentials for the Riemann and Weyl tensors.  {\it Gen. Rel. 
Grav.}, {\bf 15}, 375.

3. Illge, R. (1988).  On potentials for several classes of spinor and tensor fields in curved space-times.  {\it Gen. Rel. Grav.}, {\bf 20}, 
551.
 
 4. Andersson, F. and Edgar S.B. (2001).   Existence of Lanczos potentials and superpotentials for the Weyl spinor/tensor.  {\it Class. Quantum Gravity}, {\bf 18}, 2297.
 
 5. Andersson, F. and Edgar S.B. (2001).  Local existenece of symmetric spinor potentials for symmetric (3,1)-spinors in Einstein space-times. {\it J. Geom. and Physics}, {\bf 37}, 273-290.

 6.  Senovilla, J. M. M.  (2000).  Super-energy Tensors.   {\it Class. Quantum Grav.}, {\bf 17}, 2799.

7. Edgar, S.B. and H\"oglund A.  (2000).  The Lanczos potential for Weyl candidates exists only in four dimensions. 
 {\it Gen. Rel. Grav.}, {\bf 32},  2307.

8. Lovelock, D. (1970).   Dimensionally dependent identities.  {\it Proc. Camb. Phil. Soc.},{\bf 68}, 345-350.

9. Edgar, S.B. and H\"oglund, A.  (2002). Dimensionally dependent tensor
identities by double antisymmetrisation.  {\it  J. Math. Phys.}, {\bf 43}, 659-677.

10. Edgar, S.B. and Wingbrant, O.  (2003).  Old and new results for superenergy tensors using dimensionally
dependent identities.  {\it  J. Math. Phys.}, {\bf 44}, 6140-6159.

11. Edgar, S.B. (2004).  On the structure of the new electromagnetic conservation laws. 
{\it Class. Quantum Grav.}, {\bf  21},  L21-L25.

 12. Edgar, S.B. (2004).  Necessary and sufficient conditions for $n$-dimensional conformal Einstein spaces via dimensionally dependent identities.   Preprint: arXiv:math.DG/0404238

13. Edgar, S. B.  (1994).   The wave equation for the Lanczos tensor/spinor and a new tensor identity.  {\it Mod. Phys. Lett. A}, {\bf 9}, 479.
 
14. Edgar, S. B. and H\"oglund, A. (1997).  The Lanczos potential for the Weyl curvature tensor: existence, wave equations and algorithms.  {\it Proc. Roy. Soc. A}, {\bf 453}, 835.

15.  Atkins, W. K. and Davis, W. R. (1980).  Lanczos-type potentials in non-Abelian gauge theories. {\it Il Nuovo Cimento}, {\bf 59 B}, 116-132.

16. Roberts, M. D. (1989). Dimensional reduction and the Lanczos tensor.  {\it Mod. Phys. Letts. A}, {\bf 4}, 2739-2746.

17. Friedlander, F. G. (1975). {\it The Wave Equation on a Curved Spacetime}, Cambridge University Press.

 \end